\documentclass[a4paper]{spie}
\usepackage[]{graphicx}
\usepackage{amsmath,amssymb}
\renewcommand{\vec}[1]{\boldsymbol{{\bf #1}}}

\title{Exact analytical expression for the electromagnetic
field in a focused laser beam or pulse}

\author{Alexander M. Fedotov, Konstantin Yu. Korolev and
Maxim V. Legkov \skiplinehalf  Moscow Engineering Physics
Institute,\\ 115409 Kashirskoe sh., 31, Moscow, Russia}

\authorinfo{Corresponding author: A.M.F., E-mail: fedotov@cea.ru}

\begin{document}
\maketitle

\begin{abstract}
We present a new class of exact nonsingular solutions for the
Maxwell equations in vacuum, which describe the electromagnetic
field of the counterpropagating focused laser beams and the
subperiod focused laser pulse. These solutions are derived by the
use of a modification of the "complex source method", investigated
and visualized.
\end{abstract}

\keywords{analytical solutions, Maxwell equations, focused laser
pulse, complex source method}

\section{INTRODUCTION}
\label{sec:intro}

According to predictions of QED, vacuum can be modified by strong
electromagnetic fields if the strength becomes greater than
$F_0=m^2c^3/e\hbar=1.32\cdot 10^{16}V/cm$ and, in particular, it
should behave as a nonlinear media with respect to transmission of
such fields. Different aspects of this phenomenon (including,
e.g., vacuum polarization, elastic photon-photon scattering,
vacuum birefringence, photon splitting, pair production, etc.) had
been investigated theoretically very actively, especially after
the beginning of sixties. However, at that time the researchers
restricted themselves by considerations of the simplest
configurations of the electromagnetic field, namely, a constant
homogeneous field and a plane wave field. It was motivated,
partially, by simplicity (the Dirac equation in these backgrounds
admits exact solutions). But also, it was commonly doubted that
the fields with the strength of the order of $F_0$ will be really
created in a laboratory in viewable times. However, it was
declared recently that this experimental task could be handled on
the basis of the modern level of
technology\cite{Shen89,Mourou02,Bulanov03} and moreover,
resolution was announced to happen already in the coming
decade\cite{Mourou02}. The high-intensity field, which can be
obtained by implementation of the proposals cited above, will be
most probably realized in ultrashort (maybe, subperiod) tightly
focused laser pulses of optical range, though alternative projects
based on XFEL are also under consideration. Therefore, in order to
plan experiments in a near future, one should, firstly, examine
whether previous calculations of quantum effects in a strong field
can be applied to more realistic fields, and secondly, look for
those new effects, which can arise in realistic fields but vanish
identically in the homogeneous or the plane wave fields.

In view of this, study of nonlinear QED effects has being risen
essentially. Let us mention here just a few of these recent
studies, which are most closely related to consideration of the
realistic structure of the electromagnetic fields expected in
future experiments (for a review of other popular experimental
proposals see\cite{Marklund06}). In the
papers~\cite{Narozhny00,Narozhny02,Keitel02} scattering of
relativistic electrons on a focused laser beams and pulses was
considered. In particular, in \cite{Narozhny00,Narozhny02} the
asymmetry of the scattered electrons observed on the experiments
was explained. The papers~\cite{Bulanov04,Narozhny04} dealt with
pair creation from vacuum by a super intense ($E\sim F_0$) focused
laser pulse or by two colliding pulses. In
Refs.~\cite{Fedotov07,Narozhny07} a new effect of odd harmonics
generation by a tightly focused laser beam in vacuum was
investigated. Particularly, it was shown
in~Refs.\cite{Bulanov04,Fedotov07,Narozhny07}, that the vacuum
quantum effects induced by a tightly focused laser beam or pulse
can become observable when the peak strength of the field will
reach the values of magnitude of about one order less than the
characteristic value $F_0$. Another message of these papers was
that typically these effects are crucially sensitive to the
parameters and the particular type of the model of the field.

However, consideration of the studies mentioned above was based on
the approximate model of the focused laser field, which was
suggested in Ref.~\cite{Narozhny00} (consideration of
Ref.~\cite{Keitel02} was based on a similar but quite different
model which was suggested independently by K. McDonald though not
published). This model can be considered as a generalization of
the well-known paraxial approximation for a Gaussian beam to the
case of a transverse vector field (note, that in
Ref.~\cite{Narozhny00}~, among the other issues, the
classification of polarization types of the focused beam was
discussed exhaustively). To take into account finiteness of
temporal duration of the pulse, one, roughly speaking, multiplies
the carrier field by the slowly varying temporal envelop. It is
obvious that this approach is valid only for sufficiently long and
weakly focused pulses. In order to check the validity of these
approximations, it is desirable to have exact focused beam and
pulse-like solutions of the Maxwell equations in vacuum.

Perhaps, the most elegant attempt to construct exact analytical
beam or pulse-like field models was made in
Refs.~\cite{Deschamps71,Einziger87,Heynman87,Heynman89,Ziolkowski89}
for a scalar field and was recently directly generalized to the
case of transverse vector fields in Refs.~\cite{Wang03,Becker06}.
The main idea of this technique (now called in the literature "the
complex source method") is to generate the focused fields in
vacuum by the pointlike sources whose position is shifted into the
complex domain (this is mathematically equivalent to analytical
extension of a retarded potential). A first exact analytical model
of the focused electromagnetic pulse obtained by this method was
manifested in Ref.~\cite{Becker06}~. Being compared with the
approximations described above, this model revealed two new
qualitative features. First, the time average of the electric
field at any fixed spatial point vanishes identically, which is
very reasonable. And second, it possesses some temporal analog of
the Gouy phase which provides chirping at the wings of the pulse.
This last effect was called in ~\cite{Becker06} "the self-induced
blueshift".

In this paper we show that the usual considerations of this
complex source method suffer from two disadvantages. On the one
hand, the focused field models presented in the literature thus
far\cite{Deschamps71,Einziger87,Heynman87,Heynman89,Ziolkowski89,Wang03,Becker06}
possess algebraic singularity, which can be attributed to some
real sources (unlike the fictional pointlike "complex sources")
located on a ring in the focal plane. Therefore, in a sense, these
solutions simulate propagation of the focused field through a
circular aperture bounded by the branch cut in the focal plane,
rather than the field focused in vacuum. Only in the special
limiting case of weak focusing (when the paraxial approximation
works well) this singularity escapes away to infinity and does not
affect the field in the focal spot. On the other hand, as we show
below, the complex source method provides analytical solutions
only in the form of two colliding beams or pulses (in the limit of
weak focusing one of them disappears). We also sketch the method
for separating a single pulse solution and discuss the possibility
of its representation in a simple analytical form. The
consideration below is arranged as follows. In the
Sec.~\ref{sec:NFmodel} we remind how a certain class of
electromagnetic field configurations can be parameterized by an
axially symmetric scalar field configuration satisfying the usual
wave equation. This transformation is very close (in fact,
coincides) to the consideration of Ref.~\cite{Narozhny00}.
However, in order to increase the clarity of presentation we use
different notations and reasoning. Existence of this mapping
between the scalar focused fields and a wide class of transverse
vector focused fields with clear polarization type (say,
circularly polarized e- or h-fields) prevents us from
consideration of the more complicated vector version of the
"complex source method"\cite{Wang03,Becker06}. Therefore, in
Sec.~\ref{sec:c-source} we restrict ourselves to the more clear
scalar version of the complex source method and demonstrate its
drawbacks. The modifications which resolve these problems and the
nonsingular beam- and pulse-like solutions of the Maxwell
equations are presented in Sec.~\ref{sec:solutions}. Our
conclusions are collected in Sec.~\ref{sec:concl}.

\section{The Narozhny-Fofanov model}
\label{sec:NFmodel}

Consider a laser beam or pulse focused along the $z$-axis. In the
three-dimensional transverse gauge, the electromagnetic field is
expressed through the vector potential $\vec{A}$, which satisfies
equations\footnote{We use such units that $c=\hbar=1$.}
\begin{equation}\label{Maxwell}
\ddot{\vec{A}}-\triangle\vec{A}=0,\quad \nabla\cdot\vec{A}=0.
\end{equation}
Let $\rho$, $\phi$ and $z$ be the cylindrical coordinates. Since
the pulse is confined in radial direction, the electric and
magnetic fields are generally not orthogonal to the direction of
propagation of the focused beam or pulse. However, as usually, an
arbitrary field can be represented as a superposition of the
e-field ($\vec{E}$ orthogonal to $z$) and the h-field ($\vec{H}$
orthogonal to $z$). Since an h-field can be always obtained from
the e-field by a duality transformation $\vec{E}\to\vec{H}$,
$\vec{H}\to-\vec{E}$, it is enough to consider, say, the case of
the e-field. Hence, in the following we assume $A_z=0$. In the
simplest case of circular polarized field, the pulse consists of
photons with $j_z=+1$, where $\vec{j}$ is the total angular
momentum of the photon (assembled from the orbital momentum
$\vec{l}$ and the spin $\vec{s}$). This means, that
\begin{equation}\label{j_eigenstate}
\frac1{i}\frac{\partial\vec{A}}{\partial\phi}+i\,\hat{\vec{z}}\times\vec{A}
=\vec{A},
\end{equation}
where $\hat{\vec{z}}$ is the unit vector in $z$-direction (compare
to discussion of spherical waves in Refs.\cite{Hietler,LL4}). The
general solution to (\ref{j_eigenstate}) can be written in the
form
\begin{equation}\label{j_eigenstate_sol}
\vec{A}(\rho,\phi,z,t)=\vec{e}_+F_1(\rho,z,t)-\vec{e}_-F_2(\rho,z,t)e^{2i\phi},
\end{equation}
where $\vec{e}_\pm=\hat{\vec{x}}\pm i\hat{\vec{y}}$, and $F_{1,2}$
-- the arbitrary functions (evidently, the latter equation
provides decomposition of a state with $j_z=l_z+s_z=+1$ onto the
states with $l_z=0,\;s_z=+1$ and with $l_z=+2,\;s_z=-1$). By
substitution of the anzatz (\ref{j_eigenstate_sol}) into the
equations (\ref{j_eigenstate}) one obtains the following equations
for the auxiliary functions $F_{1,2}$,
\begin{eqnarray}\label{F1eq}
\triangle F_1=\frac1{\rho}(\rho
F_{1\rho})_\rho+F_{1zz}=\ddot{F_1},\\
\label{F2eq} \left(\triangle-\frac4{\rho^2}F_2\right)=\ddot{F_2},\\
\label{F1F2} F_{1\rho}=F_{2\rho}+\frac2{\rho}F_2.
\end{eqnarray}
The electric and magnetic fields can be expressed in terms of
functions $F_{1,2}$ by
\begin{eqnarray}\label{Ef}
\vec{E}=-\dot{\vec{A}}=-\left(\vec{e}_+F_{1t}-\vec{e}_-F_{2t}e^{2i\phi}\right),\\
\label{Hf}
\vec{H}=\nabla\times\vec{A}=-i\left(\vec{e}_+F_{1z}+\vec{e}_-F_{2z}e^{2i\phi}
-2\,\hat{\vec{z}}F_{1\rho}e^{i\phi}\right).
\end{eqnarray}
The equations (\ref{F1eq}), (\ref{F2eq}) and (\ref{F1F2}) are
compatible and $F_2$ can be expressed through $F_1$ by
\begin{equation}\label{F1F2rel}
F_2=F_1-\frac{2}{\rho^2}\int\limits_0^\rho F_1 \rho\, d\rho.
\end{equation}
However, it is more suitable to express both functions $F_{1,2}$
in terms of a unique auxiliary function $U(r,z,t)$ through
\begin{equation}\label{U-def}
F_{1,2}=U_{\rho\rho}\pm\frac1{\rho}U_\rho,
\end{equation}
then the equation (\ref{F1F2}) is satisfied identically. The new
function $U$ obeys just the scalar wave equation
\begin{equation}\label{U-eq}
\ddot{U}-\triangle U=0,
\end{equation}
and it is easy to check that the vector potential can be expressed
in the following compact form
\begin{equation}\label{A_U}
\vec{A}=2[\vec{e}_+\cdot(\hat{\vec{z}}\times\nabla)](\hat{\vec{z}}\times\nabla
U).
\end{equation}
For example, by choosing an approximate monochromatic solution
(i.e., applying a paraxial approximation)
\begin{equation}\label{NFmodel}
U=-\frac{iE_0L}{4\omega^2(1+2i\chi)}\exp\left(-i\varphi-\frac{\xi^2}{1+2i\chi}\right),
\end{equation}
for the wave equation (\ref{U-eq}), where $E_0$ is the field
strength amplitude, $\xi=\rho/R$, $\chi=z/L$, $R$-- the focal
radius ($R\gg\lambda=2\pi/\omega$), $L=\omega R^2$-- the
diffractive length, and $\varphi=\omega(t-z)$-- the fast phase,
one comes to the particular focused beam field model, which was
previously proposed by Narozhny and Fofanov and successfully
applied to the problems of electron
scattering~\cite{Narozhny00,Narozhny02}, pair
creation~\cite{Bulanov04,Narozhny04} and harmonics
generation~\cite{Fedotov07,Narozhny07} by a focused laser field.

In this way it follows that in order to construct a realistic,
vector model of the field in a focused electromagnetic beam or
pulse, it is sufficient to consider a cylindrically symmetric
scalar model first, and then apply the mapping (\ref{A_U}). In the
next section, let us consider the original complex source method
for derivation of scalar focused beam (pulse)-like solutions of
the Eq.~(\ref{U-eq}).

\section{The complex source method}
\label{sec:c-source}

The idea of the complex source method looks as follows. First,
consider a scalar wave equation with an arbitrary variable
pointlike source $q(t)$ located at the origin,
\begin{equation}\label{U_rhs}
\ddot{U}-\triangle U=4\pi q(t)\delta(\vec{r}).
\end{equation}
Obviously, the created field is expressed in terms of the usual
retarded potential,
\begin{equation}\label{U_sol1}
U(\vec{r},t)=\int\limits_{-\infty}^{+\infty}D^{(ret)}(\vec{r},t-t')q(t')\,dt'=
\frac{q(t-r)}{r},
\end{equation}
where $D^{(ret)}(\vec{r},t)$ is a retarded Green function for the
wave equation.

Now, let us shift the location of the source into the complex
domain of the $z$ axis, $\vec{r}\to\vec{r}-ib\hat{\vec{z}}$ where
$b>0$. It is also suitable to accompany this by a temporal shift
$t\to t-ib$. Then, the analytical continuation of the retarded
potential (\ref{U_sol1}) reads
\begin{equation}\label{U_sol2}
U(\vec{r},t)=\frac{q(t-ib-{\cal R})}{{\cal R}},\quad {\cal
R}^2=\rho^2+(z-ib)^2.
\end{equation}
In addition, let us choose the branch of the square root in the
definition of ${\cal R}$ so that ${\rm Im}({\cal R})\le 0$, then
in terms of the standard root we have ${\cal
R}=(z-ib)\sqrt{1+\rho^2/(z-ib)^2}$. Then, according to the
arguments presented in
Refs.~\cite{Deschamps71,Einziger87,Heynman87,Heynman89,Ziolkowski89}~,
the real part of the field (\ref{U_sol2}) describes a scalar field
focused along the $z$ axis and, in particular, satisfies the
source free wave equation (\ref{U-eq}). The viewable meaning of
this complex shift transformation can be better understood by
consideration in the $\vec{k}$-space. The equation (\ref{U_sol1})
can be represented in the form
\begin{equation}\label{U_sol3}
U_{\vec{k}\omega}=-\frac{4\pi}{(\omega+i0)^2-k^2}q_\omega,
\end{equation}
and the shift transformation is reduced to $U_{\vec{k}\omega}\to
\exp[b(k_z-\omega)]U_{\vec{k}\omega}$. On the mass shell, we have
$\omega=k$ so that
\begin{equation}\label{c_shift}
U_{\vec{k}\omega}\to
\exp[-b\omega(1-\cos{\theta})]U_{\vec{k}\omega}\approx
\exp(-b\omega\theta^2/2)U_{\vec{k}\omega},\quad b\gg\lambda
\end{equation}
where $\theta$ denotes the angle between ${\bf k}$ and the $z$
axis.

Thus, the complex shift suppresses those wave vectors in Fourier
decomposition of the radiated field which are highly inclined with
respect to $z$, as it is indeed expected for a focused field. For
example, if one chose $q(t)=q_0e^{-i\omega t}$ then, in the limit
$b\gg\lambda$, $\rho\lesssim (b\lambda)^{1/2}\ll
(b^3\lambda)^{1/4}$ by applying the expansion ${\cal R}\approx
z-ib+\rho^2/2(z-ib)$, one can reduce (\ref{U_sol2}) to the form
(\ref{NFmodel}) with $L=2b$ and $E_0=-8\omega^2q_0/L^2$. On merely
these grounds it was proposed in the literature to introduce an
additional envelop factor in $q(t)$ and to consider the equation
(\ref{U_sol2}) as an exact focused pulse-like solution of the wave
equation with the shift parameter $b$ identified with the half of
the diffractive length.

Nevertheless, the above arguments are not convincing enough for
the following two reasons. First of all, an attempt to consider
the Eq.~(\ref{U_sol2}) as a solution for the homogeneous wave
equation is incorrect and is based on the illusion that the source
term in the RHS of the Eq.~(\ref{U_rhs}) after the shift, being
proportional to $\delta(\vec{r}-ib\hat{\vec{z}})$, should vanish
identically when the vector $\vec{r}$ possess real valued
components. In fact, a $\delta$--function of a complex argument
behaves in a more complicated way than its real argument analog.
Presence of the sources follows already from a simple observation
that the transformation (\ref{c_shift}) does not replace the pole
singularities in the expression (\ref{U_sol3}) at the mass shell
by $\delta(\omega^2-k^2)$. The location of the sources coincides
to the ring singularity $\rho=b$, $z=0$ of the expression
(\ref{U_sol2})\footnote{Strictly speaking, due to untypical choice
of the branch of a square root for ${\cal R}$, the sources are
present in the exterior of this ring at the focal plane, which is
the branch cut, as well.}. This ring is picked out physically by
zero delay time for radiation emitted by the complex source and
crossing the physical hypersurface ${\rm Im}(z)=0$. Moreover, when
approaching this singularity along the focal plane $z=0$ we have
${\cal R}\propto \sqrt{|\rho-b|}$, so that $U\propto
|\rho-b|^{-1/2}$. This growth is faster than $\log[1/|\rho-b|]$,
which could be expected if the linear density of sources was
finite at the ring. Since the focal radius $R\sim
(b\lambda)^{1/2}$, the singularity becomes safe due to its escape
sufficiently far away from the focal region which contains the
most part of the field energy only in the weak focusing limit
$b\gg\lambda$. However, the paraxial approximation works well
exactly in this limit, so that knowledge of the exact solution
becomes unnecessary in this case.

Another drawback of the Eq.~(\ref{U_sol2}) can be seen from
Eq.~(\ref{c_shift}). Indeed, although due to a complex shift the
wave vectors with $\theta\gg (b\omega)^{-1/2}$ become suppressed
relatively to those with $\theta\lesssim(b\omega)^{-1/2}$,
nevertheless they remain be present. The factor $\exp[b(k-k_z)]$
just mimics but can not replace the true factor $\theta(k_z)$
which is expected to enter the Fourier decomposition of a single
beam or pulse focused along $z$. Therefore, the Eq.~(\ref{U_sol2})
actually describes two counterpropagating focused fields of
non-commensurable amplitudes. Only in the limit $b\gg\lambda$ the
amplitude of a pulse propagating in the direction opposite to $z$
becomes exponentially suppressed.

\section{The nonsingular beam and pulse-like solutions}
\label{sec:solutions}

The most obvious and direct way to correct the Eq.~(\ref{U_sol2})
and obtain the true, nonsingular solution for Eq.~(\ref{U-eq}) is
to take a difference of the retarded and advanced solutions, i.e.
to start from
\begin{equation}\label{U_sol1_m}
U(\vec{r},t)=\int\limits_{-\infty}^{+\infty}\left[D^{(ret)}(\vec{r},t-t')
-D^{(adv)}(\vec{r},t-t')\right]q(t')\,dt'=
\frac{q(t-r)-q(t+r)}{r},
\end{equation}
instead of (\ref{U_sol1}). Being a difference between two
solutions of the same inhomogeneous equation, the field
(\ref{U_sol1_m}) undoubtedly represents a solution for the
corresponding homogeneous, i.e., the wave equation. In particular,
(\ref{U_sol1_m}) is non-singular at the origin (if $q$ is
differentiable, of course) and its Fourier transform
$$
U_{\vec{k}\omega}=-4\pi\left[\frac1{(\omega+i0)^2-k^2}-\frac1{(\omega-i0)^2-k^2}\right]
q_\omega=8\pi^2i\,{\rm sgn}(\omega)\delta(\omega^2-k^2)q_\omega
$$
vanishes everywhere except on the mass shell.

Applying the complex shift transformation of the previous section
to solution (\ref{U_sol1_m}), one obtains
\begin{equation}\label{U_sol2_m}
U(\vec{r},t)=\frac{q(t-ib-{\cal R})-q(t-ib+{\cal R})}{{\cal
R}},\quad {\cal R}=\sqrt{\rho^2+(z-ib)^2},
\end{equation}
(since (\ref{U_sol1_m}) is even with respect to $r$, an arbitrary
branch of the square root defining ${\cal R}$ can be chosen in
this case).

In order to construct a monochromatic focused beam-like solution,
let us choose $q(t)=q_0e^{-i\omega t}$, where $q_0$ is a
normalization constant. Then we have
\begin{equation}\label{two_beams}
U(\vec{r},t)=2iq_0\exp[-\omega b-i\omega t]\frac{\sin(\omega{\cal
R})}{{\cal R}}.
\end{equation}
Particularly, in paraxial approximation [$b\gg\lambda$,
$\rho\lesssim (b\lambda)^{1/2}\ll (b^3\lambda)^{1/4}$], by
expanding ${\cal R}\backsimeq z-ib+\rho^2/2(z-ib)$, we obtain
$$
U(\vec{r},t)\backsimeq
\frac{iq_0}{b(1+iz/b)}\left\{\exp\left[i\omega(z-t)-\frac{\omega\rho^2}{2b(1+iz/b)}\right]-
\exp\left[-2\omega
b-i\omega(z+t)+\frac{\omega\rho^2}{2b(1+iz/b)}\right]\right\}.
$$
The second term in the parentheses (which had arisen due to a
modification proposed here) is exponentially suppressed since
$\omega^{-1},\,\rho\ll b$. Therefore, by comparison with
Eq.~(\ref{NFmodel}), we can identify the parameter $b$ with a half
of the diffractive length $L$ and $q_0$ with $-E_0L^2/8\omega^2$.
In the opposite (formal) case of ultra tight focusing
$b\ll\lambda$, our solution (\ref{two_beams}) reduces to a
superposition of smooth convergent and divergent spherical waves,
unlike (\ref{U_sol2}), which becomes a convergent spherical wave
in the half space $z<0$ and a divergent wave in the half space
$z>0$.

The auxiliary functions $F_{1,2}$ introduced in
Sec.~\ref{sec:NFmodel}, in the case under consideration read
\begin{eqnarray}\label{F1_beam}
F_1=\frac{2iq_0\exp[-\omega b-i\omega t]}{{\cal R}^5}
\left\{\left[2{\cal R}^2-3\rho^2\right]\left[\omega{\cal
R}\cos(\omega{\cal R})-\sin(\omega{\cal
R})\right]-\omega^2\rho^2{\cal R}^2\sin(\omega{\cal R})\right\},\\
\label{F2_beam} F_2=-\frac{2iq_0\exp[-\omega b-i\omega t]}{{\cal
R}^5} \left\{3\omega{\cal R}\cos(\omega{\cal
R})+\left(\omega^2{\cal R}^2-3\right)\sin(\omega{\cal R})\right\},
\end{eqnarray}
and the components of the vector potential and the electric and
magnetic fields can be obtained by plugging the expressions
(\ref{F1_beam}), (\ref{F2_beam}) into the formulas
Eq.~(\ref{j_eigenstate_sol}), (\ref{Ef}) and (\ref{Hf}) and taking
the required derivatives.

As well as the ordinary complex source method, up to now our
modification always generates pairs of counterpropagating focused
fields. In order to separate out a single focused beam, one can
multiply the Fourier transform of the field by the factor
$\theta(\pm k_z)$, which is equivalent to a transformation $U\to
U_\pm$,
\begin{equation}\label{single_pulse_tr}
U_{\pm}(\rho,z,t)=\pm\frac1{2\pi
i}\int\limits_{-\infty}^{+\infty}\frac{U(\rho,z',t)}{z'-z\mp
i0}\,dz'.
\end{equation}
The integral in (\ref{single_pulse_tr}) can be evaluated by
enclosing the integration contour and computation of residues if
$U$ vanishes at infinite $z$, e.g., if it is given by a rational
function. In this particular case, the transformation
(\ref{single_pulse_tr}) projects $U$ onto a space of functions
which are analytical in the upper half of the complex plane $z$.

Let us discuss in brief application of this transformation to the
beam-like solution (\ref{two_beams}). For this purpose, let us
first represent it in a more viewable explicit form
\begin{equation}\label{two-beam-struct}
U\propto
\frac{\exp\left[i\omega(z-ib)\sqrt{1+\frac{\rho^2}{(z-ib)^2}}\,\right]-
\exp\left[-i\omega(z-ib)\sqrt{1+\frac{\rho^2}{(z-ib)^2}}\,\right]}
{(z-ib)\sqrt{1+\frac{\rho^2}{(z-ib)^2}}}.
\end{equation}
For definiteness, in what follows consider extraction of $U_+$.
After substitution into the formula (\ref{single_pulse_tr}) one
obtains two integrals over $z'$, each one corresponding to one's
exponent on the RHS of Eq.~(\ref{two-beam-struct}). The first
integral can be enclosed in the upper half plane, while the second
integral -- in the lower one. Furthermore, in the case $\rho<b$
the first integrand in the upper half plane possesses a pole at
$z'=z+i0$ and the two branch points at $z'=i(b\pm\rho)$, while the
second one is analytical in the lower half plane. Consequently,
$U_+$ equals to a contribution in $U$ proportional to the first
exponent, added by an additional contribution given by the first
integral taken over a closed contour passing around the branch cut
connecting the branch points. This correction, however, can not be
represented in terms of elementary functions. Similarly, in the
case $\rho>b$ the first integrand possesses a pole at $z'=z+i0$
and a branch point at $z'=i(b+\rho)$, while the second integrand
possesses a unique singularity at $z'=-i(\rho-b)$. In this case,
the branch cuts go from the branch points to imaginary infinity
and do not intersect the real axis. Thus, the correction to the
contribution in $U$ proportional to the first exponent is given by
two integrals over the banks of these cuts. Finally, at the ring
$\rho=b$ the second branch point approaches the real axis and
$U_+$ becomes singular. Therefore, the nonsingular solution
(\ref{two_beams}) is in fact a superposition of two singular
counterpropagating beams.

\begin{figure}
\begin{center}
\begin{tabular}{cc}
\includegraphics[height=6cm]{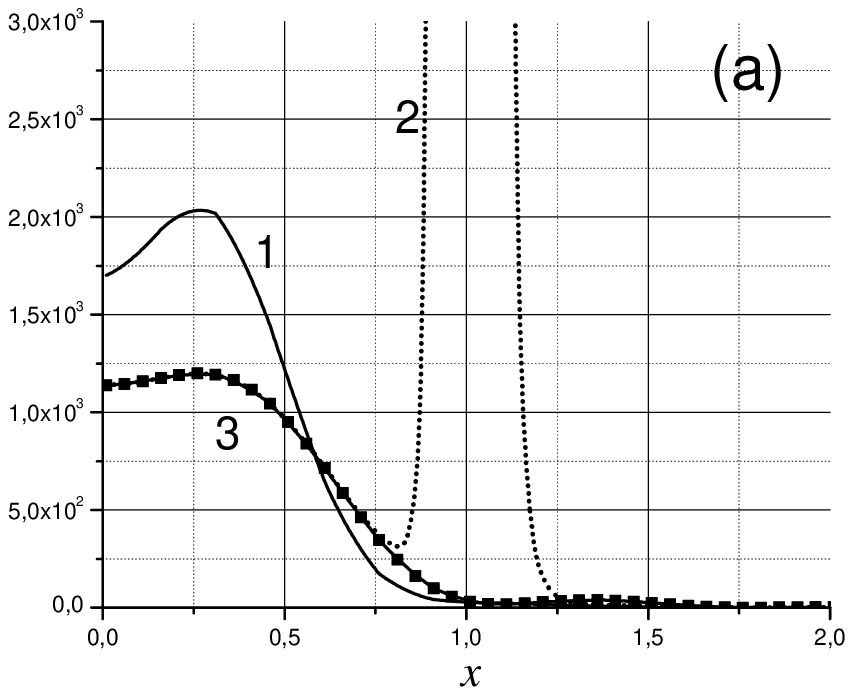}&
\includegraphics[height=6cm]{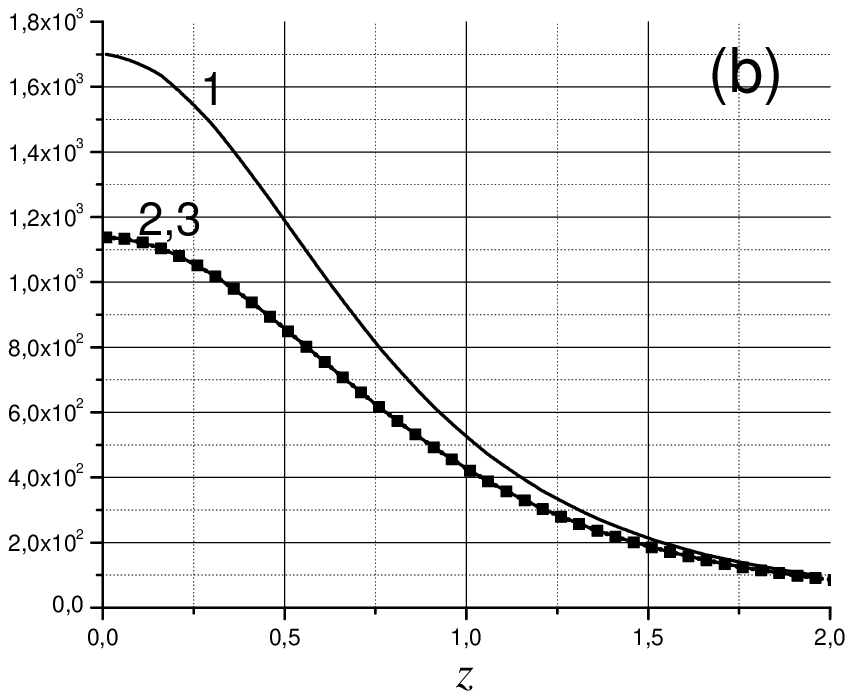}\\
\includegraphics[height=6cm]{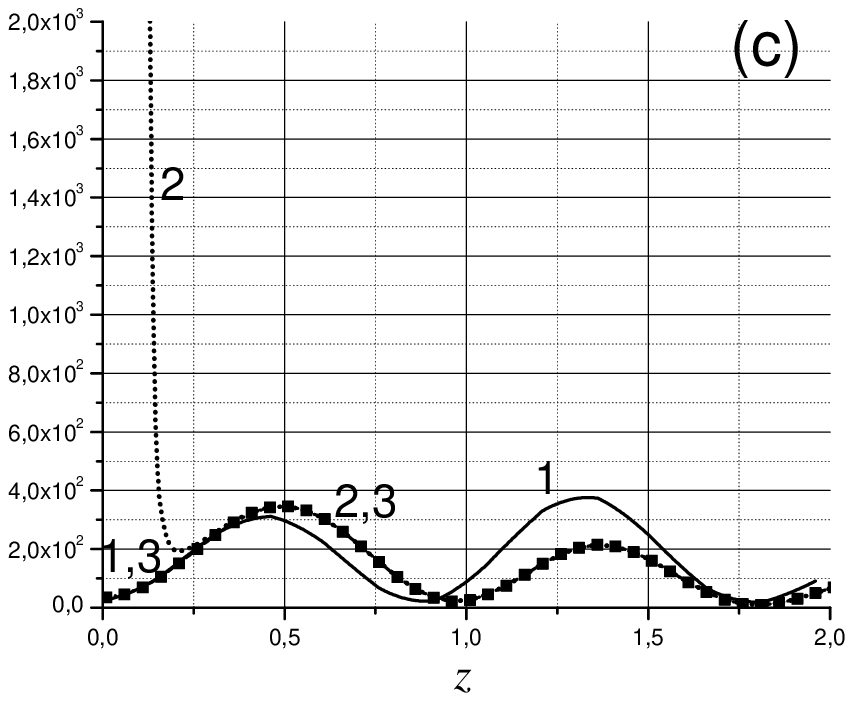}&
\mbox{\parbox{8cm}{\vspace{-6cm}\caption{\label{fig1} The spatial
distribution of the electromagnetic energy density in the
monochromatic focused beam: the modified complex source method
(curve 1), the complex source method (curve 2) and the paraxial
approximation (the Narozhny-Fofanov model, curve 3). Panel (a):
Distribution in the focal plane ($z=\phi=0$), panel (b):
distribution along the focal axis, and panel (c): distribution
along the line parallel to the focal axis and crossing the ring
singularity. The values of the parameters used are $b=1$, $q_0=1$,
$\omega=5$ ($\lambda=1.26$).}}}
\end{tabular}
\end{center}
\end{figure}

The spatial distributions of the electromagnetic energy density
\begin{equation}\label{en_dens}
w=\frac1{8\pi}\left\{\left[{\rm Re}\,(\vec{E})\right]^2+\left[{\rm
Re}\,(\vec{H})\right]^2\right\},
\end{equation}
in the three models of a monochromatic focused beam [the paraxial
approximation defined by Eq.~(\ref{NFmodel}), the ordinary complex
source method defined by Eq.~(\ref{U_sol2}), and the modified
complex source method defined by Eq.~(\ref{two_beams})] are
compared at the Fig.~\ref{fig1} in the tight focusing regime
($b\sim\lambda$). It is clear from the figure, that in this regime
the field model corresponding to the ordinary complex source
method occurs to be very close to paraxial approximation (the
Narozhny-Fofanov model) everywhere but in the vicinity of the ring
singularity, where it breaks down. At the same time, the solution
obtained by the modified complex source method, being exact and
nonsingular, differs from both of them.

The same technique can be applied to construction of the
pulse-like solutions as well. In order to do this, it is enough to
introduce an envelop factor in $q(t)$, e.g., of the Gaussian form
$e^{-t^2/2T^2}$ or of the Lorentzian form $(1+t^2/T^2)^{-1}$,
where $T$ is the duration parameter of the pulse. As it is clear
from the consideration above, the result is an expression for the
field components in terms of elementary functions, though
generally a rather complicated one. Below, let us consider a
special but perhaps the most interesting case of an ultrashort,
subperiod focused pulse. An especially simple model of a pulse can
be obtained if one drops the carrier frequency factor in $q(t)$
totally. Furthermore, since for real values of $t$, $z$ and $\rho$
we have $|{\rm Im}({\cal R})|\le b$ and hence $-2b\le{\rm
Im}(t-ib\pm{\cal R})\le 0$, in order to obtain a nonsingular
solution for the whole range of the duration parameter, let us
choose $q(t)$ to be analytic in the lower half plane of complex
$t$. Assuming $q(t)=q_0/(1+it/T)$ and applying (\ref{U_sol2_m}),
we have
\begin{equation}\label{short_pulses}
U(\vec{r},t)=\frac{2iq_0T}{(T+b+it)^2+{\cal R}^2}.
\end{equation}
The  nonsingular solutions corresponding to single pulses
propagating along (opposite to) $z$-axis can be extracted from
Eq.~(\ref{short_pulses}) by using (\ref{single_pulse_tr}) and are
given by
\begin{equation}\label{short_pulse}
U_\pm(\vec{r},t)=\mp\frac{q_0T}{\sqrt{(T+b+it)^2+\rho^2}\,[z-ib\pm
i\sqrt{(T+b+it)^2+\rho^2}]},
\end{equation}
respectively. Thus, unlike the beam-like solution considered
above, our solution (\ref{short_pulses}) represents a
superposition of two nonsingular counterpropagating focused
fields. A more general nonsingular solution of this kind can be
constructed as an arbitrary superposition
$c_1U_+(z,\rho,t)+c_2U_+(z,\rho,-t)$ with complex coefficients.
Note also, that in the weak focusing limit $b\to\infty$, $T={\rm
const}$, and both expressions (\ref{short_pulses}) and
(\ref{short_pulse}) reduce to a plane wave moving along $z$ with
the profile $U\backsimeq iq(t-z)/b$.

\begin{figure}
\begin{center}
\begin{tabular}{c}
\includegraphics[height=9cm]{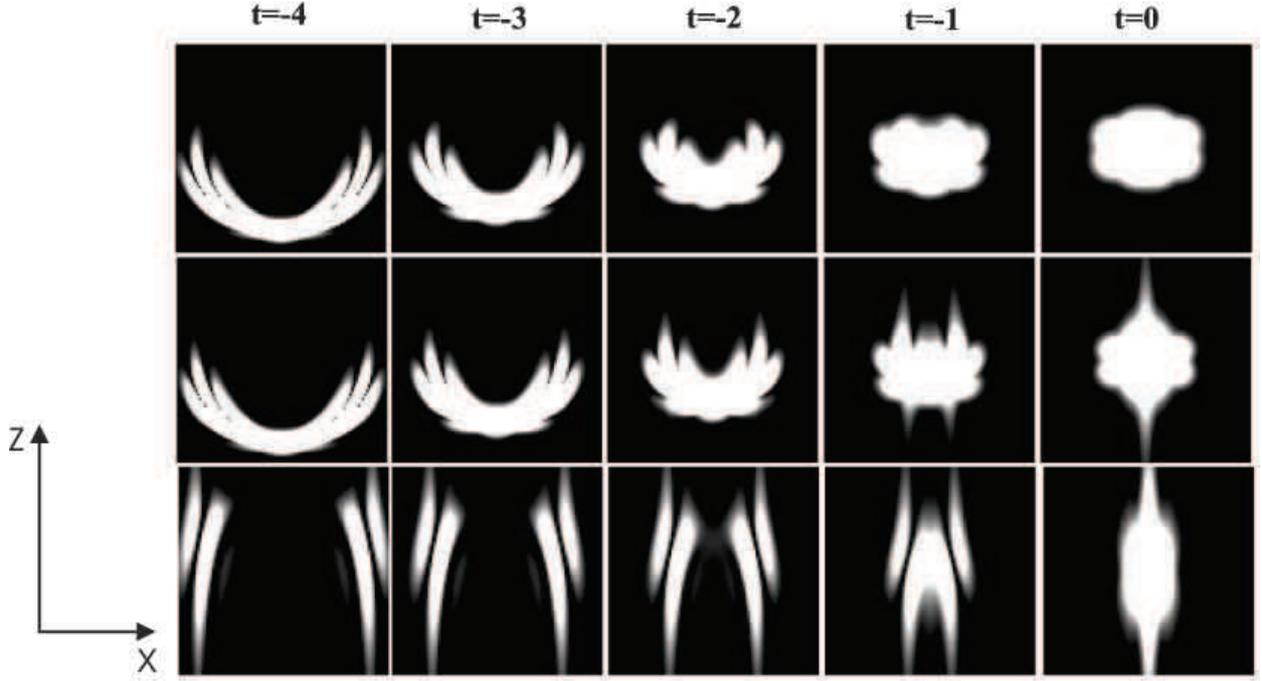}
\end{tabular}
\end{center}
\caption{\label{fig2} The distribution of the electromagnetic
energy density in the $xz$-plane in subsequent stages of focusing
of the subperiod pulse-like solutions corresponding to $U$ (the
top row), $U_+$ (the middle row) and $U_-$ (the bottom row). The
values of the parameters used are $b=1$, $T=0.2$.}
\end{figure}

The resulting expressions for the vector potential and the
electromagnetic fields are quite cumbersome and this short paper
is not an appropriate place to adduce them. Rather, let us present
a visualization of the electromagnetic energy distribution in the
$xz$-plane, corresponding to Eqs.~(\ref{short_pulses}) and
(\ref{short_pulse}), at different stages of focusing, see
Fig.~\ref{fig2}. It can be seen from the pictures, that even in
the tight focusing regime $U_+$ actually almost coincides to $U$
(i.e., $U_-$ is negligible) except for the moments which are
directly close to the moment of maximal collapse. Also, it can be
recognized that unlike $U$ and $U_+$, the backward propagating
pulse $U_-$ possesses a coreless tubular structure.

In conclusion of the section, let us present an independent
derivation of the solution (\ref{short_pulses}) on the basis of
direct evaluation of the corresponding Fourier integral. According
to the beginning of the section, the Fourier transform to start
with is given by
\begin{equation}\label{Ukw}
U_{\vec{k}\omega}=16\pi^3iq_0T\theta(\omega)e^{bk_z-\omega
(T+b)}\delta(\omega^2-k_\perp^2-k_z^2).
\end{equation}
Let us evaluate the Fourier integral $U(\vec{r},t)= \int
d\omega\,d^3k\,e^{i(\vec{k}\cdot\vec{r}-\omega
t)}U_{\vec{k}\omega}/(2\pi)^4$ using the cylindrical coordinates
in the $k$-space. After performing the integration with respect to
the azimuthal angle (which is trivial) and the integration with
respect to $k_z$ (which is realized by taking away a
$\delta$-function), we are left with
\begin{equation}\label{Fint1}
U(\vec{r},t)=2iq_0T\int\limits_0^{+\infty}d\omega\,e^{-i\omega(t-ib-iT)}
\int\limits_0^\omega
\frac{dk_\perp\,k_\perp}{\sqrt{\omega^2-k_\perp^2}}\,J_0(k_\perp\rho)
\,\cos\left[(z-ib)\sqrt{\omega^2-k_\perp^2}\right].
\end{equation}
The inner integral over $k_\perp$ equals\cite{RG}
$\sin(\omega{\cal R})/{\cal R}$. Hence, the final integral over
$\omega$ is elementary and we come to the formula
(\ref{short_pulses}). Note that in order to reproduce the
Eq.~(\ref{short_pulse}) in the same way one should take the
integral which is similar to the inner one in (\ref{Fint1}), but
with the imaginary exponent instead of the cosine. However, we
could not find it in the mathematical literature thus far.

\section{Conclusions}\label{sec:concl}

In this paper, we have explicitly demonstrated  that the "complex
source method"
~\cite{Deschamps71,Einziger87,Heynman87,Heynman89,Ziolkowski89,Wang03,Becker06}
of generating the exact focused beam or pulse-like solutions of
the wave or Maxwell equations should be modified in order to
obtain the fields which are nonsingular in the focal plane,
correspondingly to absence of the sources. Also, we have shown
that this technique always generates solutions describing a pair
of counterpropagating focused fields, rather than a single one. We
suggest resolution to both disadvantages. In order to generate
nonsingular solutions, one should exploit the difference between
the retarded and advanced potentials instead of choosing a special
branch of the retarded potential, as it was previously accepted in
the literature. Besides, we propose the regular method which
allows one to separate the resulting field onto the one
propagating along the focal axis and the backward propagating one.
However, even if one starts from the nonsingular solution, it is
not guaranteed generally that the forward and backward propagating
fields will remain nonsingular. Our final proposal is to start
with an appropriate anzats for the vector potential in terms of
scalar functions obeying the wave equation, and then to construct
them by application of the complex source method. This allows one
to control the polarization of the electromagnetic field
beforehand. A reasonable example of such anzats with known
polarization type (e- or h- circularly polarized wave) was
proposed in Refs.~\cite{Narozhny00, Narozhny02}.

As an application of the modified "complex source method", we
obtained and visualized the solution which describes a
counterpropagating focused monochromatic laser beams, and the
solution describing a subperiod focused laser pulse. However, the
technique under consideration is in no way restricted to the
examples given and allows one to obtain a wide class of focused
beam- or pulse-like fields. An advantage of this approach is that
the field components can be expressed in terms of elementary
functions, though generally in a rather complicated way.
Nevertheless, it seems that such expressions are very suitable for
numerical calculations of both the motion of the charged particles
interacting with the laser field and the quantum effects induced
by it, especially in the tight focusing regime. The key point for
such applications of our expressions is absence of singularities
in them.

\acknowledgments

A.M.F. is grateful to W.~Becker who attracted our attention to the
"complex-source method". We are also grateful to S.~P. Goreslavki,
S.~R. Kelner and especially to N.~B. Narozhny for valuable
discussions and comments. This work was supported in part by the
Russian Foundation for Basic Research (grant 06-02-17370-a), the
Ministry of Science and Education of Russian Federation and the
Russian Federation President grants MK-2364.2007.2,
NSh-320.2006.2.

\end{document}